\documentclass[fleqn,usenatbib]{mnras}

\usepackage{newtxtext,newtxmath}

\usepackage[T1]{fontenc}
\usepackage{ae,aecompl}

\usepackage{graphicx}	
\usepackage{amsmath}	
\usepackage{amssymb}	

\newcommand{\grb}{GRB\,100205A}

\usepackage{soul} 
\usepackage{xargs} 
\usepackage[pdftex,dvipsnames]{xcolor}  
\usepackage{amsmath}
\title[GRB100205A at High-Redshift]{The Case for a High-Redshift Origin of GRB\,100205A}

\author[A. A. Chrimes et al.]{A. A. Chrimes,$^{1}$\thanks{E-mail: A.Chrimes@warwick.ac.uk}
A. J. Levan,$^{1}$
E. R. Stanway,$^{1}$
E. Berger,$^{2}$
J. S. Bloom,$^{3}$ \newauthor
S. B. Cenko,$^{4,5}$
B. E. Cobb,$^{6}$
A. Cucchiara,$^{7}$
A. S. Fruchter,$^{8}$
B. P. Gompertz,$^{1}$ \newauthor
J. Hjorth,$^{9}$
P. Jakobsson,$^{10}$
J. D. Lyman,$^{1}$
P. O'Brien,$^{11}$ 
D. A. Perley,$^{12}$ \newauthor
N. R. Tanvir,$^{11}$ 
P. J. Wheatley$^{1}$ and
K. Wiersema$^{1}$
\\
$^{1}$Department of Physics,  University of Warwick, Gibbet Hill Road, Coventry, CV4 7AL, UK\\
$^{2}$Harvard-Smithsonian Center for Astrophysics, 60 Garden Street, Cambridge, MA 02138, USA\\
$^{3}$Department of Astronomy, University of California, Berkeley, CA 94720, USA\\
$^{4}$NASA Goddard Space Flight Center, 8800 Greenbelt Road, Greenbelt, DMD 20771, USA\\
$^{5}$Joint Space-Science Institute, University of Maryland, College Park, MD 20742, USA\\
$^{6}$Department of Physics, The George Washington University, Washington, DC 20052, USA\\
$^{7}$University of the Virgin Islands, College of Science and Mathematics, \#2 Brewers Bay Road, Charlotte Amalie, USVI 00802\\
$^{8}$Space Telescope Science Institute, 3700 San Martin Drive, Baltimore, MD21218, USA\\
$^{9}$Dark Cosmology Centre, Niels Bohr Institute, University of Copenhagen, Juliane Maries Vej 30, DK-2100 Copenhagen \o, Denmark\\
$^{10}$Centre for Astrophysics and Cosmology, Science Institute, University of Iceland, Dunhagi 5, 107 Reykjavik, Iceland\\
$^{11}$Department of Physics and Astronomy, University of Leicester, University Road, Leicester, LE1 7RH, UK\\
$^{12}$Astrophysics Research Institute, Liverpool John Moores University, 146 Brownlow Hill, Liverpool, L3 5RF, UK\\
}

\date{Accepted XXX. Received YYY; in original form ZZZ}

\pubyear{2019}

\begin{document}
\label{firstpage}
\pagerange{\pageref{firstpage}--\pageref{lastpage}}
\maketitle

\begin{abstract}
The number of long gamma-ray bursts (GRBs) known to have occurred in the distant Universe ($z>5$) is small (${\sim}$15), however these events provide a powerful way of probing star formation at the onset of galaxy evolution. In this paper, we present the case for GRB\,100205A being a largely overlooked high-redshift event. While initially noted as a high-$z$ candidate, this event and its host galaxy have not been explored in detail. By combining optical and near-infrared Gemini afterglow imaging (at $t<1.3$\,days since burst) with deep late-time limits on host emission from the {\it Hubble Space Telescope}, we show that the most likely scenario is that \grb\ arose in the range $4 < z < 8$. \grb\ is an example of a burst whose afterglow, even at $\sim 1$ hour post burst, could only be identified by 8-m class IR observations, and suggests that such observations of all optically dark bursts may be necessary to significantly enhance the number of high-redshift GRBs known. 
\end{abstract}

\begin{keywords}
gamma-ray burst: individual: 100205A -- galaxies: high redshift
\end{keywords}



\section{Introduction}
Long-duration gamma-ray bursts (GRBs) give rise to a synchrotron afterglow, detectable at optical wavelengths if sufficiently rapid and deep follow-up observations are made. A substantial fraction, however, lack such emission even when it would expected from extrapolation of the X-ray spectral slope \citep{1998ApJ...502L.123G,2001A&A...369..373F}. When the X-ray to optical spectral slope, ${\beta}_{\mathrm{OX}}$, is below the recognised threshold of 0.5, the event is classified as `dark' \citep{2004ApJ...617L..21J}. This is typically evaluated at 11 hours post-burst to avoid contamination from early-time effects including X-ray flares and plateaus. An alternative method uses ${\beta}_{\mathrm{OX}} < {\beta}_{\mathrm{X}} - 0.5$ to define darkness \citep{2009ApJ...699.1087V}. There are two primary causes for darkness in GRBs: attenuation by dust, or rest frame ultraviolet H\textsc{I} absorption at high redshift \citep[e.g.,][]{1999ApJ...512L...1F,levan06,2009AJ....138.1690P,2011A&A...526A..30G,svensson12,perley13,2013ApJ...767..161Z,2019MNRAS.484.5245H,2019MNRAS.486.3105C}. The number of GRBs known at high-redshift ($z>5$, in the epoch of reionisation) is small \citep[$\sim$15, from around 500 GRBs with a known or estimated redshift, ][]{2006Natur.440..184K,2006GCN..5155....1C,2006GCN..4545....1G,2006AIPC..836..552J,2007ApJ...669....1R,Salvaterra,2009ApJ...693.1610G,2009Natur.461.1254T,2011ApJ...736....7C,2011A&A...526A.154A,2013GCN.14796....1C,2014ApJ...781....1L,2014GCN.15936....1J,2014GCN.16269....1C,Tanvir_hi-z}, and each one is valuable, as they provide insight into star formation in the low mass, low luminosity galaxies which power the epoch of reionisation. Because they have small projected offsets from their hosts, high-redshift GRBs with a detected afterglow uniquely allow us to place accurate, deep upper limits on the luminosities of the faintest, undetected galaxies, probing fainter galaxies than deep field studies \citep{2007ApJ...665..102B,2012ApJ...754...46T,2012ApJ...749L..38T,ManyHiZ}. For those with the brightest afterglows, insight into the burst environment can be gained from absorption lines in their spectra (e.g. \citealt{2006Natur.440..184K,2014arXiv1405.7400C,sparre14,2015A&A...580A.139H}).

In this paper, we present the case for dark GRB\,100205A being a high-redshift event, undetected in the $r$-band, but faintly visible in the infrared, suggestive of the presence of the Lyman-$\alpha$ break between the $r$ and $J$ bands at a redshift $z>5$.

\section{Observations, Data Reduction and Results}
\grb\ ($T_{90}=26$\,s) was detected by the {\it Neil Gehrels Swift Observatory} \citep{Swift} on 5 Feb 2010 \citep{2010GCN.10361....1R}. The Burst Alert Telescope \citep[BAT,][]{BAT} measured a fluence of ($4.0\pm0.7\times10^{-7}$)\,erg\,cm$^{-2}$, with a peak photon flux of ($0.4\pm0.1$)\,cm$^{-2}$\,s$^{-1}$ (15-150\,keV, 90 per cent confidence errors). The enhanced X-Ray Telescope \citep[XRT; ][]{XRT} position was ra.\,09h 25m 33.08s, dec.\,31$^{\circ}$ 44$^{\prime}$ 24.3$\arcsec$, with a 90 per cent error radius of 1.7${\arcsec}$ \citep{2009MNRAS.397.1177E} \footnote{\url{http://www.swift.ac.uk/xrt_positions/}}.

The X-ray afterglow was rapidly identified, and ground based observations were taken in the first hour after the burst. However, none of these early optical observations revealed a candidate optical 
afterglow \citep[see][]{2010GCN.10362....1M,2010GCN.10365....1C,2010GCN.10366....1T,2010GCN.10374....1C,2010GCN.10383....1N,2010GCN.10375....1U,2010GCN.10399....1P}, marking \grb\ as a dark burst \citep{2010GCN.10362....1M}, and motivating further follow-up.

\subsection{Gemini}
Gemini/GMOS-S \citep{2004PASP..116..425H} observations in the $r$-band were obtained 40 minutes post trigger. These observations were reduced in the standard fashion
within the Gemini {\sc IRAF} environment, and did not yield an optical afterglow to a 3\,$\sigma$ limit of $R>25.2$, the deepest upper limit on the optical light available. 

Given this non-detection the burst location was subsequently imaged
in the infrared by Gemini-N/NIRI \citep{2003PASP..115.1388H} in the $Y$, $J$, $H$ and $K$ bands starting at ${\sim}$2.4 hours post-burst, as shown in Table \ref{tab:observations}. 

\begin{figure}
  \centering
  \begin{minipage}[b]{0.9\columnwidth}
    \hspace*{0\textwidth}
    \includegraphics[width=0.9\columnwidth]{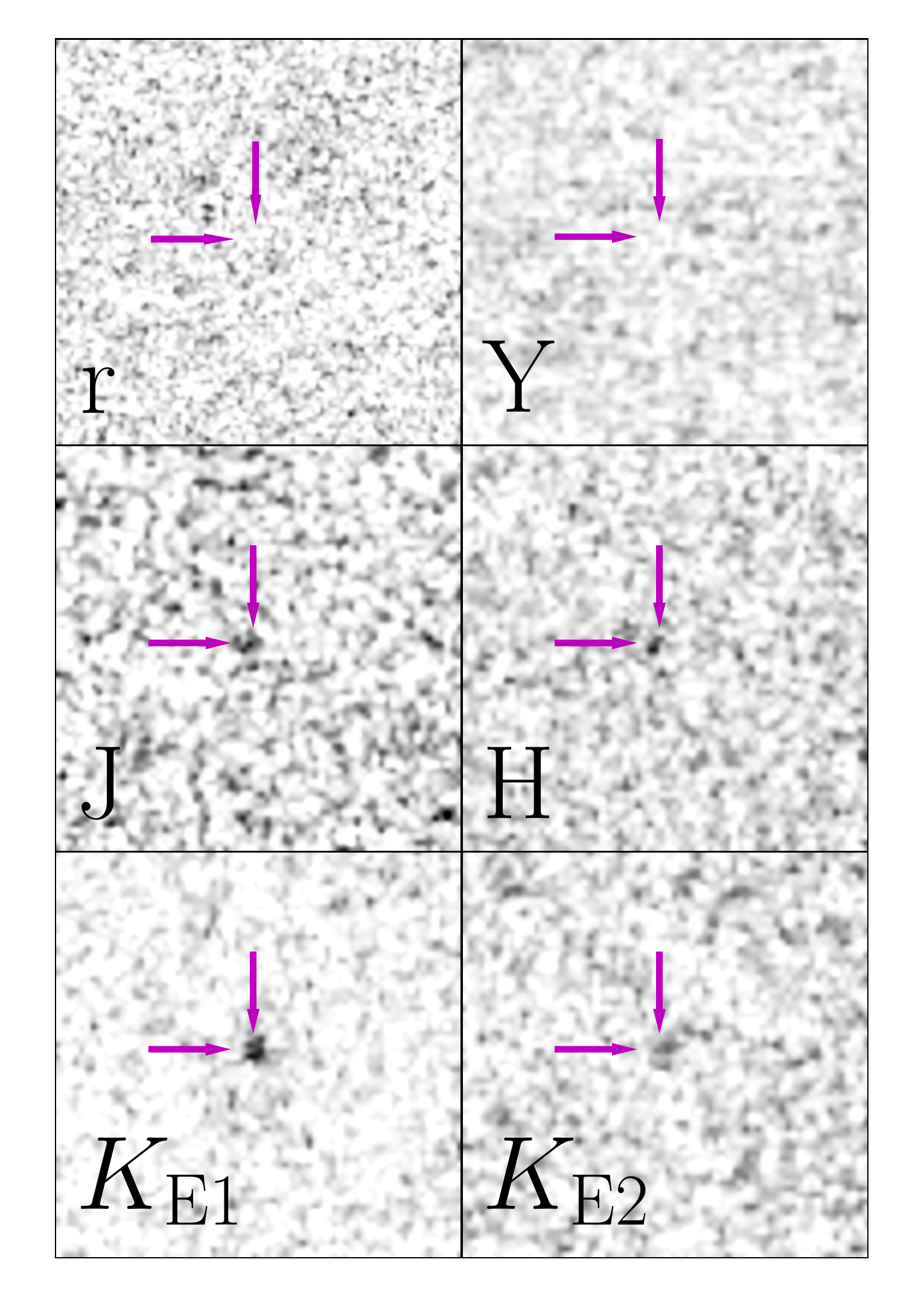}
  \end{minipage}
  \vfill
  \begin{minipage}[b]{0.88\columnwidth}
    \hspace*{0\textwidth}
    \includegraphics[width=0.88\columnwidth]{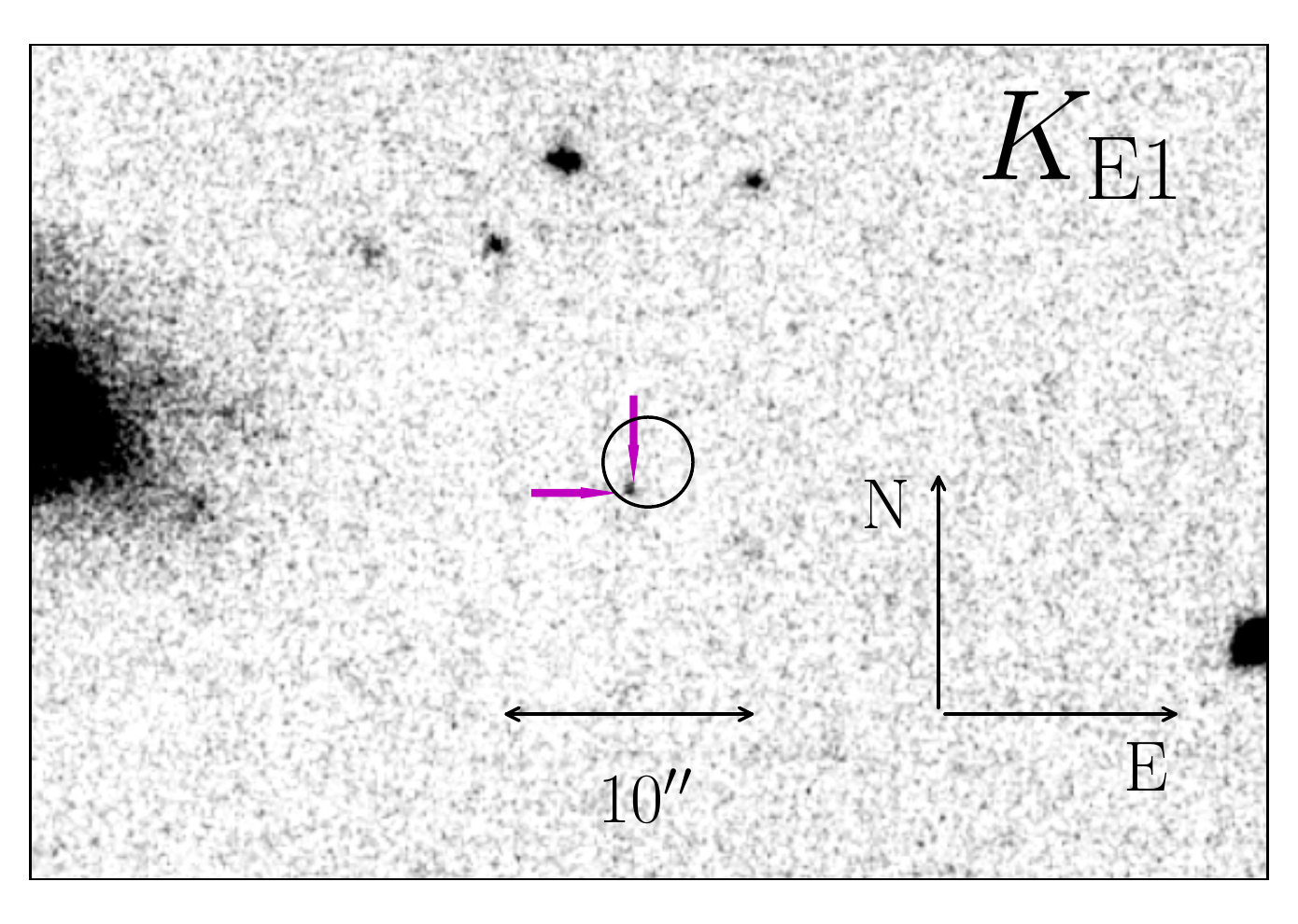}
    \caption{Image cutouts ($4\times4\arcsec$) around the afterglow position for each set of Gemini observations. Numbering indicates whether data is from the first or second observation with that filter. Below these is wider area cutout to demonstrate the location of the burst with respect to a nearby large galaxy. Included here is the {\it Swift} enhanced XRT position, indicated by a black circle (with a 90 per cent error radius of 1.7$\arcsec$). All images have been smoothed with a $3{\times}3$ pixel Gaussian filter.}
    \label{fig:gemini}
  \end{minipage}
\end{figure}

The data were reduced using standard procedures with the Gemini {\sc IRAF} package, and care was taken to optimise bad pixel rejection. Cutouts of the reduced images around the GRB afterglow location are shown in Figure \ref{fig:gemini}. Also shown is a wider-field view, which includes the {\it Swift} enhanced XRT position. The $K$-band numbering ($K_\mathrm{E1}$, $K_\mathrm{E2}$) refers to the first and second epochs of observation, which were approximately 1 day apart. The two epochs in $H$ and $J$ are sufficiently close in time that we have combined the data from these, where that led to an improvement in signal to noise. Afterglow aperture magnitudes are listed in Table \ref{tab:observations}. The photometric aperture radii are equal to the FWHM for each image, and background subtraction was performed using annuli around these apertures. The aperture positions were anchored to the same point, relative to field objects, in each image. The $K_\mathrm{E1}$ afterglow centroid was used as the reference position. Photometry was calibrated against UKIDSS \citep{2007MNRAS.379.1599L} in $J$, $H$ and $K$, and Pan-STARRS \citep{2016arXiv161205560C} in $r$ and $Y$.
All magnitudes are given in the AB system and are corrected for foreground Galactic dust extinction using a $R_{V}=3.1$ Fitzpatrick reddening law \citep{1999PASP..111...63F} and the dust maps of \citet{2011ApJ...737..103S}, with E(B-V)=0.0165. An afterglow was securely detected in the $J$, $H$ and $K$-band images, at a location consistent with that of the X-ray afterglow detected by {\it Swift}/XRT.

\begin{table}
\centering 
\caption{Gemini/GMOS-S ($r$-band) and Gemini/NIRI afterglow observations. Afterglow magnitudes (or 3\,${\sigma}$ limits), corrected for Galactic extinction, and detection significance are listed in the final two columns. T$_{\mathrm{obs}}$ is the average time since burst trigger for the input exposures.}
\begin{tabular}{l l c c c c c l} 
\hline 
Filter  & $\lambda_\mathrm{eff}$ & T$_{\mathrm{obs}}$ & N$_{\mathrm{exp}}$ & Tot. Int. &  FWHM  & Mag & ${\sigma}$ \\ 
\newline
 & [${\mu}$m] & [hrs] & & [s] & [${\arcsec}$] & \\
\hline 
$r$	&	0.62	&	0.68	&	5	&  602.5    & 1.44  & >25.2	& -	\\
$Y$	&	1.02	&	3.22	&	10	&	600    & 1.15  & >23.5	& -	\\
$K_\mathrm{E1}$	&	2.20	&	3.88	&	30	&	1800    & 0.58  & 23.45${\pm}$0.09 & 13.2		\\
$H$	&	1.63	&	4.73	&	24	&	720    & 0.65 & 23.63${\pm}$0.26 & 3.69		\\
$J$	&	1.25	&	5.48	&	17	&	510    & 0.63  & 24.29${\pm}$0.29 &	3.35	\\
$K_\mathrm{E2}$	&	2.20	&	30.32	&	28	&	1620    & 0.45  & 24.42${\pm}$0.16 & 6.40		\\
\hline 
\end{tabular}
\label{tab:observations}
\end{table}

\subsection{\textit{Hubble Space Telescope}}
The burst region was observed with Wide Field Camera 3\footnote{\url{http://www.stsci.edu/hst/wfc3}} in the F606W and F160W bands on 2010 Dec 06 (10 months post-burst, programme 11840, PI: Levan). These bands have effective wavelengths of 0.57 and 1.52$\mu$m respectively. A three-point dither pattern was observed in each band, with total integration times of 1209\,s (F160W) and 1140\,s (F606W). {\sc Astrodrizzle} (part of the {\sc drizzlepac} python package\footnote{\url{http://drizzlepac.stsci.edu}}) was used to reduce the images. The chosen {\sc pixfrac} was 0.8, with final scales of 0.065 arcsec pixel$^{-1}$ (F160W) and 0.02 arcsec pixel$^{-1}$ (F606W). 

\begin{figure}
\centering
\includegraphics[width=0.75\columnwidth]{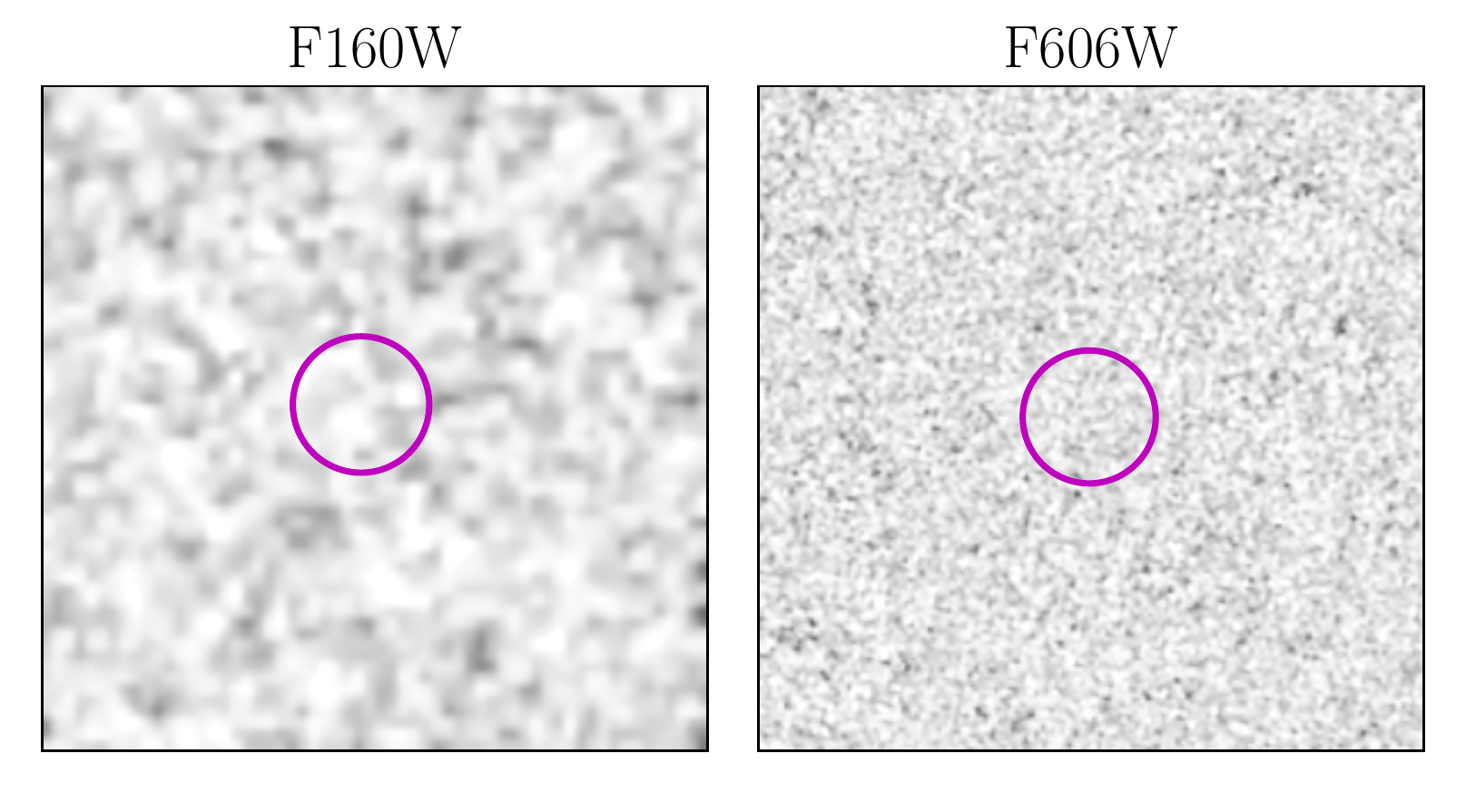}
\caption{{\it HST} image stamps at the location of GRB\,100205A, the images are smoothed with a $3{\times}3$ pixel Gaussian filter. The burst location is indicated in each image by a 0.4 arcsec radius circle. No host is detected down to 3\,${\sigma}$ magnitude limits of 26.7 and 27.1 for F160W and F606W respectively.}
\label{fig:nohost}
\end{figure}

We once again use the Gemini $K_\mathrm{E1}$ detection as a reference position, determining the burst location in the {\em HST} images by calculating a direct transformation based on six reference objects in the field. We use the {\sc IRAF} tasks {\sc geomap} and {\sc geoxytran} to fit for rotation, shifts and scaling in the $x$ and $y$ directions. The total positional uncertainty on the afterglow position in the {\it HST} frame has contributions arising from this transformation and the uncertainty on the afterglow position in the Gemini image, yielding a positional uncertainty of 18\,mas in the F160W image and 49\,mas in F606W. Figure \ref{fig:nohost} shows image cutouts centred on the burst location. The source is not detected in either band.  At the position of the afterglow, we measure 3\,${\sigma}$ magnitude limits of 26.7 in F160W and 27.1 in F606W (with a 0.4 arcsec aperture, for which STScI tabulate zero-points \footnote{\url{http://www.stsci.edu/hst/wfc3/analysis}}). A similarly deep optical limit was obtained by \citet{2010GCN.10399....1P} two days post burst, using the Low Resolution Imaging Spectrometer on Keck \citep{1995PASP..107..375O} to place an $r$-band 3\,${\sigma}$ limit of 26.7 on any host emission at the burst location.

\begin{figure*}
\centering
\includegraphics[width=0.95\textwidth]{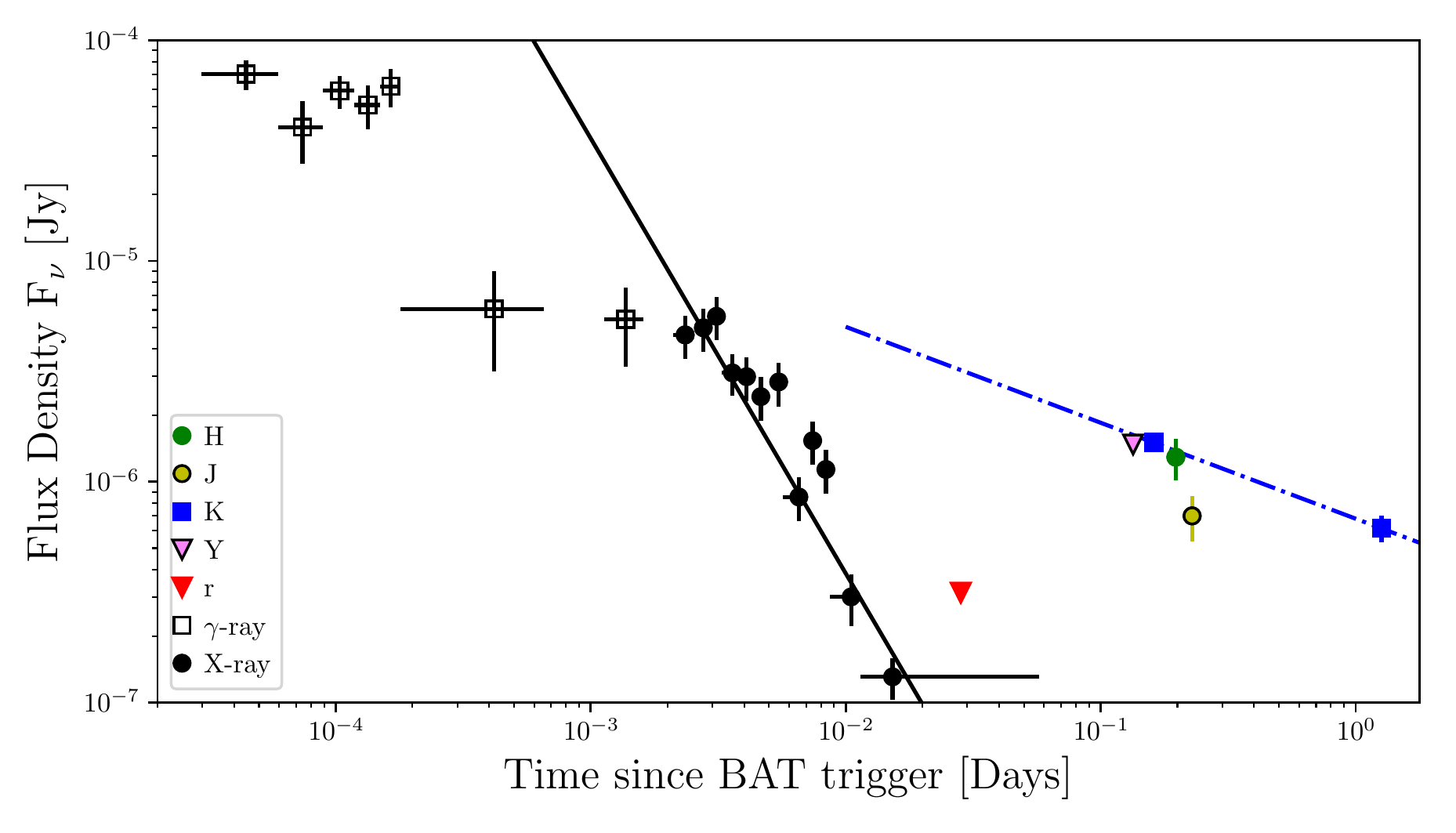}
\caption{The gamma-ray, X-ray and NIR/optical light curve of GRB\,100205A. Triangles represent 3\,${\sigma}$ upper limits, circles and squares are detections. The X-rays were undetected by {\it Swift} XRT by the time of the first optical/IR follow-up observation. The solid black line is a power law fit to the X-ray data points, representing the X-ray temporal decay, while the dashed blue line is a fit to the two $K$-band observations, giving the NIR decay rate. We do not extrapolate the NIR fit far beyond the $r$-band limit, as the prompt BAT lightcurve and NIR behaviour are likely driven by different physical mechanisms, making such a comparison misleading.}
\label{fig:lightcurve}
\end{figure*}


\section{Interpretation}
In Figure \ref{fig:lightcurve}, we show the light curve for GRB\,100250A, featuring the gamma-ray, X-ray, near-infrared (NIR) and $r$-band fluxes and limits. The prompt emission lightcurve, detected by the BAT instrument in gamma-rays, is characterised by a weak single peak with a duration of T$_{90}$ = $26.0\pm7.5$\,s. There is no evidence for continued central engine activity beyond this period, and the X-ray is not sufficiently steep to be well explained as high-latitude emission. We therefore consider the possibility that the X-ray emission arises entirely from the afterglow forward shock.
The X-ray lightcurve is monitored from a few minutes after the burst. It decays rapidly - the decay rate of $\sim t^{-2}$ at this early epoch is steeper than typically seen - becoming undetectable after about 30 minutes, before the first optical observation is made. The initial $r$-band non-detection lies chronologically between the X-ray monitoring and the start of NIR observations at about 3 hours post-burst. While the NIR data is sparse, it appears to show a less rapid decline in flux density than that seen in the X-ray. As a result, the X-ray to optical spectral energy distribution (SED) is difficult to reconstruct since there is no time overlap, and we consider two different methods for extrapolating between data points. In Figures \ref{fig:agsed} and \ref{fig:nirsed}, we construct SEDs from the afterglow measurements. The first assumes that the NIR and optical flux decays at the same rate as the X-ray, the second derives a decay rate from the two $K$-band points. We note that an extrapolation based on the prompt gamma-ray emission would lie between these, but is likely inappropriate for the late time afterglow. After considering the SED, we go on to discuss the burst energetics and the host non-detection.

\subsection{X-ray based SED construction}\label{sec:xray}
Firstly, we assume that the flux in $J$, $H$, and $K$ bands shows the same time evolution as the X-ray flux, and that the flux decays according to $F_{\nu}\propto\,t^{{\alpha}}$. All detections and the $r$-band limit are extrapolated backwards or forwards to the mean time of the first epoch of observations (0.18 days, at which point there are contemporaneous NIR observations). The X-ray temporal slope ${\alpha}=-1.97{\pm}0.14$ and 
X-ray photon spectral index ${\Gamma} = 1.91^{+0.25}_{-0.22}$ are obtained from the {\it Swift} online database\footnote{\url{http://www.swift.ac.uk/xrt_live_cat}} \citep[][90 per cent errors]{2009MNRAS.397.1177E}. The corresponding intrinsic neutral Hydrogen column density (at $z=0$) is ($3^{+6}_{-3}\times10^{20}$)\,cm$^{-2}$, a low value which disfavours a dusty, low-redshift explanation for the darkness of this GRB \citep{2010GCN.10399....1P}. 

We extrapolate the X-ray flux to the optical (NIR) using a broken power law, with the two segments of the synchrotron spectrum given by $F_{\nu}\propto\,{\nu}^{-{\Gamma}+1}$ and $F_{\nu}\propto\,{\nu}^{-{\Gamma}+1+{\Delta}{\beta}}$, where ${\Delta}{\beta}$ accounts for a synchrotron spectral cooling break between the NIR and X-ray \citep{1998ApJ...497L..17S}. ${\Delta}{\beta}=0.5$ provides a satisfactory fit in most GRBs \citep{2011A&A...526A..30G}.

The spectrum is fitted to the X-ray points, while the break frequency and break strength are allowed to vary. The parameter values which best fit the extrapolated NIR points are obtained through a procedure fully described in appendix \ref{sec:appendix}. The data are consistent either with an unbroken extrapolation (${\Delta}{\beta}=0$), or an extrapolation which breaks in the infrared (i.e. not shortwards of the $r$-band).

\begin{figure}
\centering
\includegraphics[width=0.45\textwidth]{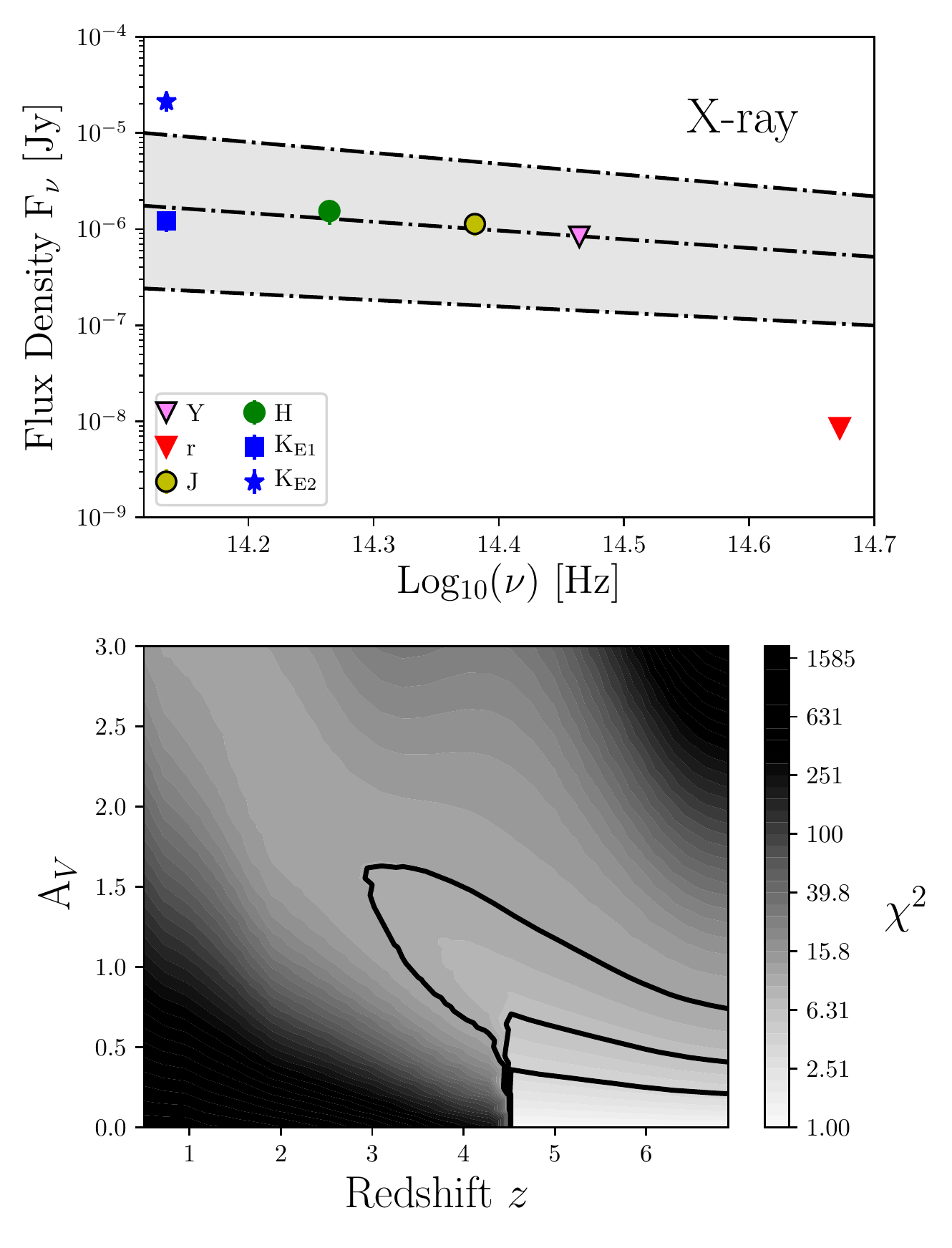}
\caption{Upper panel: the afterglow SED for GRB\,100205A, where the $J$, $H$, $K$, $Y$ and $r$-bands (triangles are 3\,${\sigma}$ upper limits) have been extrapolated to the midpoint of the first epoch of observations, assuming the same rate of dimming as measured in the X-rays. Flux uncertainties include the contribution from the uncertainty in temporal evolution. An extrapolation of the X-ray spectral slope with ${\Delta}{\beta}=0$ at the same epoch is shown, with the 90 per cent confidence region shaded and bounded by dot-dash lines. A strong break occurs between $r$ and $J$. Lower panel: ${\chi}^{2}$ minimisation over a grid of power law models. Contours representing the 67, 95 and 99.5 per cent frequentist probability intervals are overlaid in black.}
\label{fig:agsed}
\end{figure}

\begin{figure}
\centering
\includegraphics[width=0.45\textwidth]{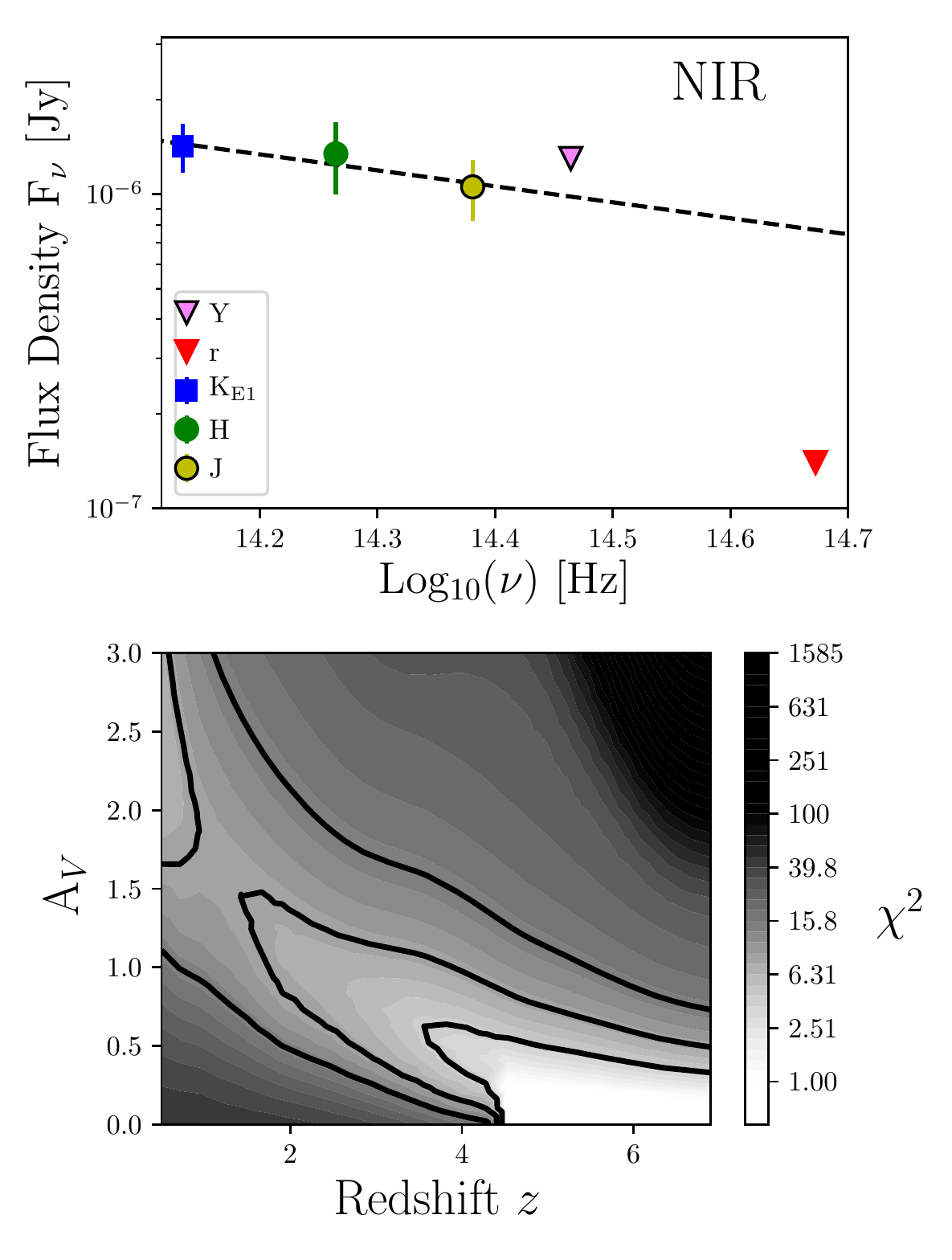}
\caption{Upper panel: As in Figure \ref{fig:agsed}, but the fluxes and limits are extrapolated using the decay rate as seen in the $K$-band. The $K_\mathrm{E2}$ point is not shown as it overlaps with $K_\mathrm{E1}$ by construction. The best fit NIR spectral slope is given by the dashed line. Lower panel: ${\chi}^{2}$ minimisation as in Figure \ref{fig:agsed}, with data extrapolated according to the $K$-band decay rate.}
\label{fig:nirsed}
\end{figure}

 The upper panel of Figure \ref{fig:agsed} illustrates the observed fluxes extrapolated in time as points with error bars and compares these against the ${\Delta}{\beta}=0$ spectral extrapolation from the X-rays. The uncertainty in the X-ray extrapolation is indicated by the shaded region, which is dominated by the uncertainty on the XRT spectral slope.
 
The NIR to X-ray spectral slope, ${\beta}_\mathrm{IR-X}$, is ${\sim}$-0.92, compared to the XRT value of -0.91 (where ${\beta} = 1-{\Gamma}$). Fit values for this interpretation of the data are listed in Table \ref{tab:fitdata}. The fit in this case is very good given the uncertainties, although we note that the extrapolated $K_\mathrm{E2}$ point is not in agreement. Host contamination in the $K$-band is effectively ruled out by the deep ${\it HST}$ non-detection in F160W, discussed in section \ref{sec:host}. Therefore, if the X-ray decay model is correct, then this epoch must have been contaminated by a flare or other non-standard variability.

Dark GRBs are typically classified based on X-ray to optical (i.e. $r$-band) rather than X-ray to NIR spectral slopes. 
\grb\ was classified as a dark burst with ${\beta}_{\mathrm{OX}}<0.28$, due to the very deep $r$-band non-detection at early times \citep{2010GCN.10362....1M}. Given a simple power law SED passing from the X-ray and through the optical limit, the NIR bands would also be expected to have a faint flux, inconsistent with the observations. In order to produce the observed $r$-band decrement relative to the X-ray to NIR fit described above, the spectrum would have to show a broken (${\Delta}{\beta}$>$0.5$) extrapolation from the X-ray to the $r$-band, followed by another sharp steepening of the slope in the narrow frequency range between $r$ and $J$ and a return to the original slope at longer wavelengths (lower frequencies) - i.e. three intrinsic spectral breaks in the afterglow. This is not consistent with any model or observation of GRB afterglow behaviour.

For the purposes of investigating the darkness of \grb\, we instead adopt a simple case where the X-ray and NIR lie on the same section of the synchrotron spectrum ($\Delta\beta=0$, or no break). We note that the best-fit broken power law from figure \ref{fig:appendix}, and this simplified model, are both consistent with the data. 

Since many dark GRBs are the result of dust extinction \citep[see for example][]{2009AJ....138.1690P,2011A&A...526A..30G,svensson12,perley13,2013ApJ...767..161Z,2014A&A...569A..93J,2015MNRAS.446.4116V,2019MNRAS.484.5245H,2019MNRAS.486.3105C}, a first assumption may be that this apparent break is in fact due to spectral curvature induced by dust absorption within the host galaxy. The precise level of the relative dust correction between the observed $r$ and $J$ bands depends on the redshift of the source. Alternatively, the break could be due to the presence of the Lyman-$\alpha$ break between the $r$ and $J$ bands \citep[e.g.][]{2006Natur.440..184K, 2009Natur.461.1254T}. 

In order to determine the likely cause of the factor 100 drop in flux to the $r$-band, we compare a grid of afterglow models to the extrapolated NIR data points and $r$-band flux (for the latter we use the 1${\sigma}$ limit, see Table \ref{tab:fitdata}). The models consist of unbroken power laws with a range of ${\beta}$ values given by the uncertainty on the XRT spectral slope. Each model is then placed at a range of redshifts ($0<z<7$), subject to a range of rest-frame dust attenuation ($0<A_{V}<3$, with an SMC-like attenuation law), and normalised to best fit the extrapolated NIR and $r$-band fluxes using ${\chi}^{2}$ minimisation. Because the $r$-band encroaches on the Lyman break from around $z{\sim}$4, we account for the filter profile \footnote{\url{http://svo2.cab.inta-csic.es/svo/theory/fps}} and include the effect of line-of-sight averaged HI absorption as a function of redshift \citep{1995ApJ...441...18M,2015ApJ...813L...8M}. The results of minimising ${\chi}^{2}$ across the grid of parameters is shown in the lower panel of Figure \ref{fig:agsed}. The $K$, $H$, $J$ and $r$-bands are used for the fitting of four variables, however these variables are not independent. We therefore conservatively assume only one degree of freedom, which defines the contours given the minimum in ${\chi}^{2}$ \citep[e.g.][]{1976ApJ...210..642A}. 

The result of this analysis is that the only region of parameter space producing acceptable fits is at high-redshift, and low dust extinctions. Because we see no evidence of the Lyman break entering the $J$ band, we use the short-wavelength edge of this band (${\lambda}{\sim}1.17{\mu}m$) to infer an upper redshift limit of ${\sim}$8. This places \grb\ in the range $4.5 < z < 8$, at the high end of the GRB redshift distribution.

Another possibility is that molecular Hydrogen, vibrationally excited by a strong ultraviolet (UV) flux, could produce absorption at rest-frame UV wavelengths \citep[shortwards of 1650\,\AA, ][]{2000ApJ...532..273D,2009ApJ...701L..63S,2018MNRAS.481.1126W}. However, molecular to atomic Hydrogen ratios are sufficiently low in GRB hosts, even when H$_{2}$ lines are detected, that it effectively rules out this scenario \citep{2019A&A...623A..43B}.


\begin{table}
	\centering
	\caption{Values of the parameters obtained from fitting power laws to the X-ray or temporally extrapolated NIR and optical data, assuming the fading rate of either the X-rays or $K$-band. Included are the temporal index ${\alpha}$ and spectral index ${\beta}$ (with 90 per cent errors), extrapolated $r$-band limits, and the flux decrement $F_\mathrm{d}$ between the observed $r$-band (for which we use the 3\,${\sigma}$ $r$-band constraint) and the model.}
	\label{tab:fitdata}
	\begin{tabular}{lp{1.4cm}p{1.4cm}cc} 
		\hline
		 & ${\alpha}$ & ${\beta}$ & Extrapolated $r$-band   & $F_\mathrm{d}$ \\
		 & & & 1 (3) ${\sigma}$ limit [Jy] &\\
		\hline
		X-ray & $-1.97\pm0.14$ & $-0.91^{+0.25}_{-0.22}$ & 2.92 (8.50) $\times10^{-9}$ & $>64$\\
		$K$-band & $-0.43\pm0.16$ & $-0.51\pm0.26$ & 0.47 (1.40) $\times10^{-7}$ & $>5.5$\\
		\hline
	\end{tabular}
\end{table}

\subsection{NIR based SED construction}
This analysis also suffers from uncertainty due to the assumed fading rate of the afterglow. For an alternative approach, we can look instead at the temporal decay of the afterglow in the $K$-band. The NIR temporal index ${\alpha}=-0.43{\pm}0.16$ (90 per cent error) is substantially different from the X-ray temporal index, warranting an alternative interpretation of the data using this decay rate instead.

In the upper panel of Figure \ref{fig:nirsed}, we extrapolate the $r$-limit and NIR fluxes to the epoch 1 mean time of 0.18 days using the $K$-band decay rate, and fit a spectral slope to the NIR points at that epoch. The best fit NIR spectral slope has the value ${\beta}_\mathrm{NIR} = -0.51\pm0.26$ (90 per cent error). The break between $r$ and $J$ is less strong in this scenario, with a flux decrement of factor ${\sim}$5. Fit parameters are listed in Table \ref{tab:fitdata}, for ease of comparison to the X-ray decay interpretation.

A single spectral break between the $J$ amd $r$ bands could, in this case, explain the photometric data. In order to do this, however, extremely low environmental densities would be required to produce such a blue break frequency at ${\sim}$4-5 hours post burst, and would be highly unusual \citep[e.g.][]{1999ApJ...523..177W,2011A&A...526A..30G}.

As for the X-ray hypothesis, we compare afterglow models to the extrapolated NIR data points and $r$-band flux in order to determine the possible cause of this spectral break. The models are once again power laws with a spread of ${\beta}$ values, bounded by the uncertainty on the NIR spectral slope. The models are subject to a range of redshifts ($0<z<7$), dust attenuations ($0<A_{V}<3$) and  normalisations. Neutral hydrogen absorption and the filter profile are accounted for as before. The lower panel of Figure \ref{fig:nirsed} shows the results of minimising ${\chi}^{2}$ over this parameter space. Although dusty and low redshift scenarios cannot be ruled out, the 67 per cent confidence region is nearly entirely limited to $z>4$ and $A_\mathrm{V}<0.5$, indicating a preference for low-dust, high-redshift solutions. The presence of emission in the $J$-band, as with the X-ray case, places an upper limit of $z<8$ on the burst, putting it in the range $4<z<8$.

We cannot rule out variability in the NIR, particularly as there are only two epochs available in the $K$-band. We note that the disagreement between the X-ray and NIR temporal slopes might indicate that there is non-afterglow activity occurring in either band. We can likely be confident that the correct decay rate and spectral slope lie somewhere between the NIR and X-ray cases. 
The non-standard X-ray afterglow argument is strengthened if the burst is indeed at high-redshift - given that the X-ray observations finished at t${\sim}$40 minutes, this corresponds to only a few minutes post burst in the rest-frame (for $4<z<8$). Such early times often show non-standard afterglow activity, including flares, the decay of which could produce the steep X-ray decline seen in this burst \citep{2006ApJ...642..389N}. 

The result that the burst lies in the range $4<z<8$ is independent of the method chosen to interpret these data as figures \ref{fig:agsed} and \ref{fig:nirsed} demonstrate.

\begin{figure}
\centering
\includegraphics[width=0.95\columnwidth]{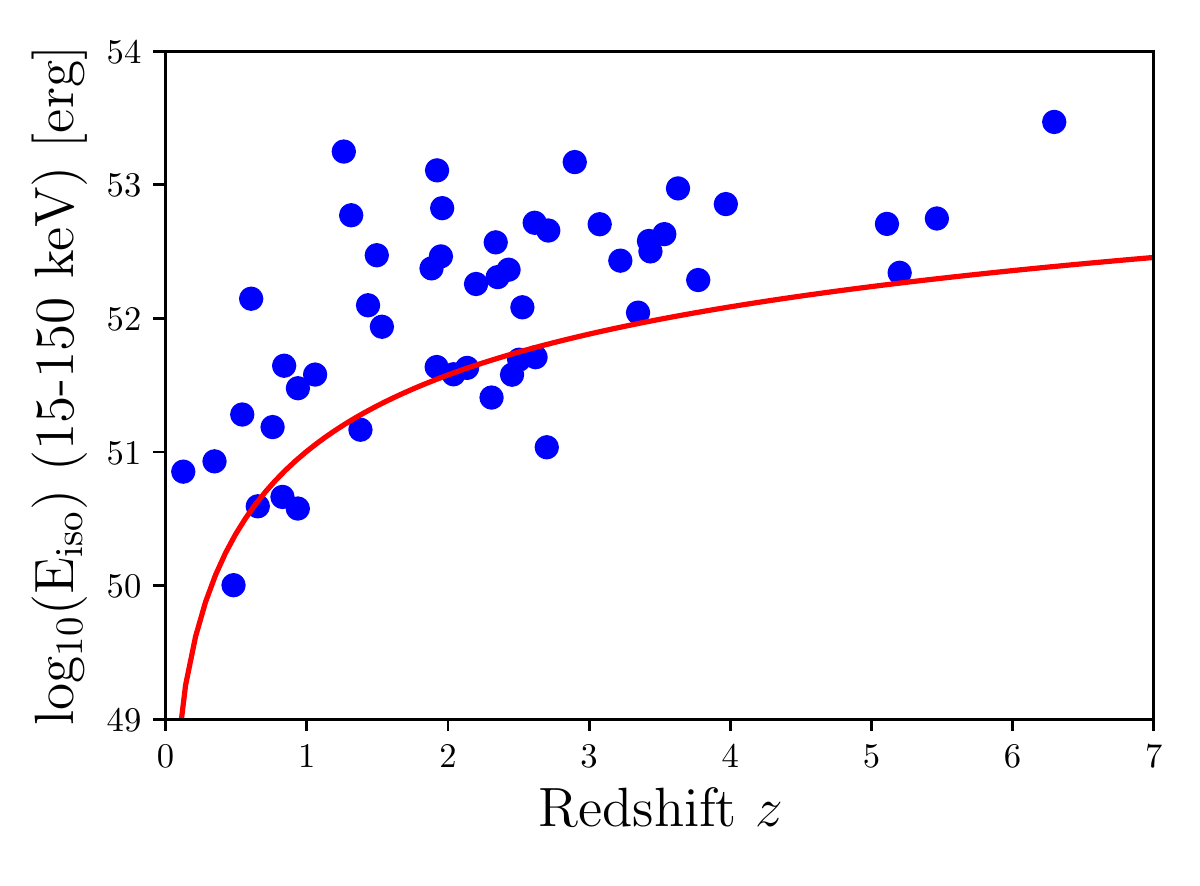}
\caption{The isotropic equivalent energy $E_{\mathrm{iso}}$ of {\it Swift} GRBs from the optically unbiased TOUGH sample versus redshift. \grb\ (indicated by the red line) is not unreasonably under-luminous at any redshift ${\gtrsim}$ 0.5, and not unreasonably luminous at any redshift. Energetics considerations therefore cannot rule out a high-redshift interpretation.}
\label{fig:eiso}
\end{figure}


\subsection{High energy properties}
The high energy properties of \grb\ can also offer some constraints. In particular, a bright burst may become a significant outlier in energetics at higher redshift, disfavouring such a distance indication. Figure \ref{fig:eiso} shows the distribution of isotropic energy inferred for {\it Swift} GRBs from the optically unbiased TOUGH sample against redshift \citep{2012ApJ...756..187H}. \grb\ is unremarkable if placed any redshift ${\gtrsim}$ 0.5, although it is at the fainter end of the luminosity distribution. The energetics of \grb\ therefore do not preclude a high-redshift interpretation.

\subsection{Non-detection of the host}\label{sec:host}
Finally, the extremely deep limit obtained for the galaxy host flux in the {\em HST} F160W band strongly favours a higher redshift origin. This is not due to the Lyman break - this feature is not redshifted into the F160W filter until $z{\sim}11-12$, and at redshifts this high no $J$ or $H$-band afterglow detection would be expected. The NIR afterglow detections in fact provide a firm upper limit on the redshift of $z\sim8$. Instead, the host non-detection implies a very low intrinsic host luminosity rather than HI absorption in the intergalactic medium. 

\begin{figure}
\centering
\includegraphics[width=0.95\columnwidth]{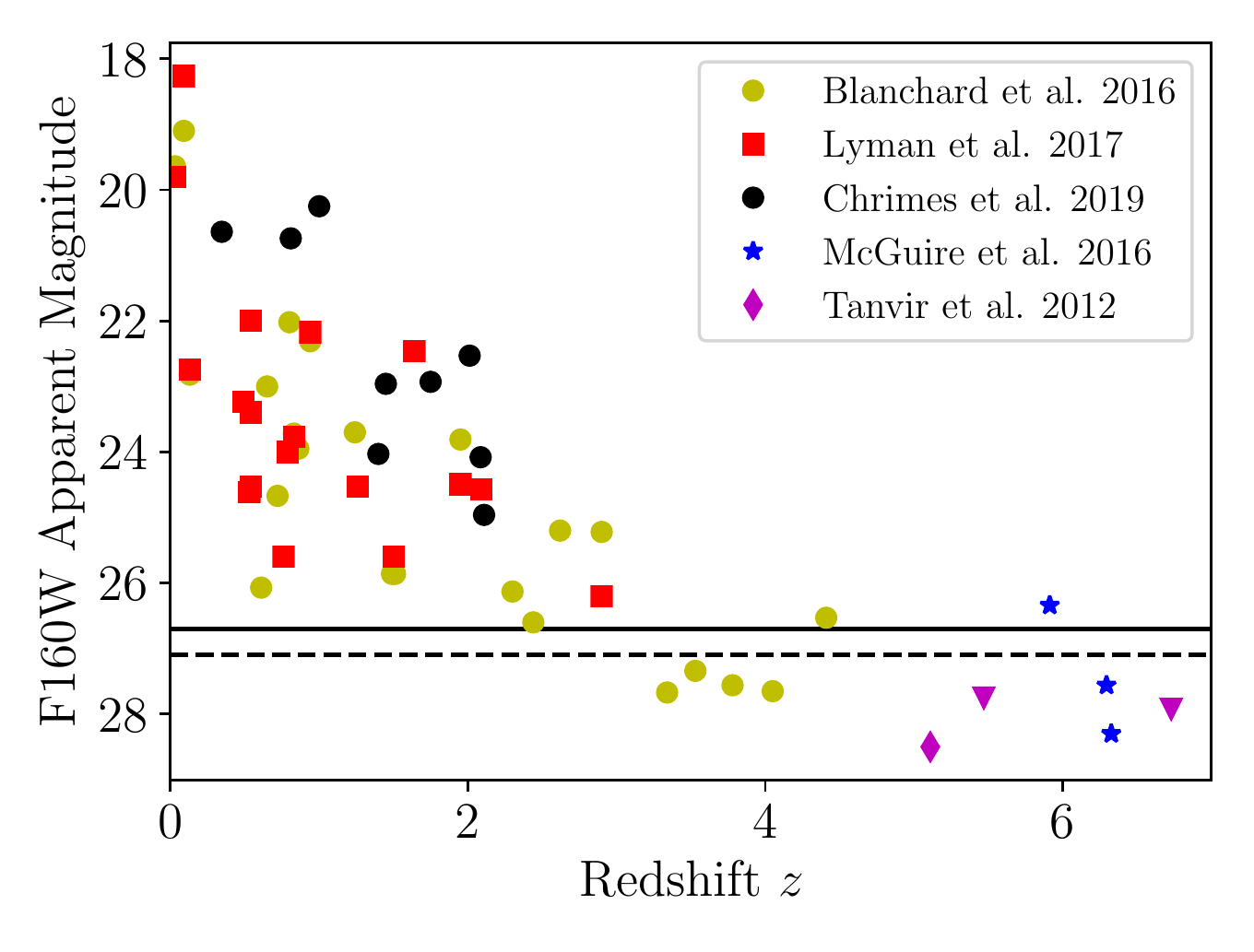}
\caption{GRB host galaxy apparent magnitudes, measured in the {\it HST} F160W band \citep[][]{2016ApJ...817..144B,2017MNRAS.467.1795L,2019MNRAS.486.3105C}. Four high-redshift GRB host detections in the $J$ and F140W bands, and two 2\,${\sigma}$ upper limits (F160W, triangles), are also shown \citep{ManyHiZ,2012ApJ...754...46T}. The 2\,${\sigma}$(3\,${\sigma}$) limit at the position of GRB\,100205A is indicated with a dashed (solid) line.}
\label{fig:appmag}
\end{figure}

In Figure \ref{fig:appmag}, we show F160W apparent magnitudes for GRB hosts with known redshift \citep[either from host emission or afterglow absorption lines,][]{2016ApJ...817..144B,2017MNRAS.467.1795L,2019MNRAS.486.3105C}. We also include three high redshift data points in F140W - GRBs\,130606A, 050904 and 140515A \citep{ManyHiZ} - in addition to one detected host (GRB\,060522, $J$-band) and two deep limits (F160W) from \citet{2012ApJ...754...46T}. The \citet{2017MNRAS.467.1795L} sample is composed exclusively of optically bright (thus $z<3$) bursts. The \citet{2019MNRAS.486.3105C} sample is composed exclusively of dark bursts. The other samples include a mixture of bursts. For redshifts $z {\lesssim} 3$, an apparent 1-2$\mu$m (observed) magnitude of $>26.7$ is uncharacteristically faint for GRB hosts, and at these lower redshifts essentially all are detected. Conversely, at $z{\gtrsim} 3$, such faint hosts become the norm, with most host galaxies undetected at this level. We note that in the sample of \citet{2019MNRAS.486.3105C}, GRB\,100205A is the {\it only} burst for which no host is detected in F160W. If we assume that \grb\ occurred at $z = 5$, the rest-frame UV absolute magnitude of the host is $M_\mathrm{UV}>-19.74$, placing it at least one magnitude fainter than M$_\ast$ at that redshift \citep{2015ApJ...803...34B} - demonstrating the ability of GRBs to select low mass star forming galaxies in the distant Universe. 

\section{Conclusions}
We have presented Gemini and {\it HST} imaging of the afterglow and host galaxy location of the dark GRB\,100205A. The lack of a detected host at $m_\mathrm{AB}\mathrm{(F160W)}>26.7$ (3\,${\sigma}$), combined with a strong spectral break in the afterglow SED between $r$ and $J$, suggests a high-redshift ($4 < z < 8$) origin for this burst, adding it to the small sample of GRBs known to have occurred in the early Universe. Despite the limited photometric coverage, this conclusion stands independent of the spectral and temporal extrapolation methods assumed. It was only identified thanks to rapid and deep optical observations that could place meaningful constraints on the darkness of a burst with an apparently faint X-ray afterglow, and subsequently inform infrared observations. This highlights that such deep observations, beyond the range of modest aperture telescopes at any epoch, may well be necessary to significantly increase the sample of known high redshift GRBs.

\section*{Acknowledgements}
AAC is supported by Science and Technology Facilities Council (STFC) grant 1763016. AAC also thanks the William Edwards educational charity. AJL, ERS and PJW have been supported by STFC consolidated grant ST/P000495/1.

Based on observations obtained at the Gemini Observatory, acquired through the Gemini Observatory Archive and processed using the Gemini {\sc IRAF} package. The Gemini Observatory is operated by the Association of Universities for Research in Astronomy, Inc., under a cooperative agreement with the NSF on behalf of the Gemini partnership: the National Science Foundation (USA), National Research Council (Canada), CONICYT (Chile), Ministerio de Ciencia, Tecnolog\'{i}a e Innovaci\'{o}n Productiva (Argentina), Minist\'{e}rio da Ci\^{e}ncia, Tecnologia e Inova\c{c}\~{a}o (Brazil), and Korea Astronomy and Space Science Institute (Republic of Korea).

Based on observations made with the NASA/ESA Hubble Space Telescope, obtained from the data archive at the Space Telescope Science Institute. STScI is operated by the Association of Universities for Research in Astronomy, Inc. under NASA contract NAS 5-26555. 

This work made use of data supplied by the UK Swift Science Data Centre at the University of Leicester.

The UKIDSS project is defined in \citet{2007MNRAS.379.1599L}. UKIDSS uses the UKIRT Wide Field Camera \citet[WFCAM;][]{2007A&A...467..777C}. The photometric system is described in \citet{2006MNRAS.367..454H}, and the calibration is described in \citet{2009MNRAS.394..675H}. The pipeline processing and science archive are described in \citet{2008MNRAS.384..637H}.

The Pan-STARRS1 Surveys (PS1) and the PS1 public science archive have been made possible through contributions by the Institute for Astronomy, the University of Hawaii, the Pan-STARRS Project Office, the Max-Planck Society and its participating institutes, the Max Planck Institute for Astronomy, Heidelberg and the Max Planck Institute for Extraterrestrial Physics, Garching, The Johns Hopkins University, Durham University, the University of Edinburgh, the Queen's University Belfast, the Harvard-Smithsonian Center for Astrophysics, the Las Cumbres Observatory Global Telescope Network Incorporated, the National Central University of Taiwan, the Space Telescope Science Institute, the National Aeronautics and Space Administration under Grant No. NNX08AR22G issued through the Planetary Science Division of the NASA Science Mission Directorate, the National Science Foundation Grant No. AST-1238877, the University of Maryland, Eotvos Lorand University (ELTE), the Los Alamos National Laboratory, and the Gordon and Betty Moore Foundation.

IRAF is distributed  by  the  National  Optical  Astronomy  Observatory,  which  is  operated  by  the  Association  of Universities for Research in Astronomy (AURA) under cooperative agreement with the National Science Foundation. This research has made use of the SVO Filter Profile Service supported from the Spanish MINECO through grant AyA2014-55216. We acknowledge the use of Ned Wright's online cosmology calculator \citep{2006PASP..118.1711W}, and Jochen Greiner's website (\url{http://www.mpe.mpg.de/~jcg/grbgen}). 

We thank the referee Bruce Gendre for their useful feedback on this paper.




\bibliographystyle{mnras}
\bibliography{grb100205A} 



\appendix
\section{Fitting the X-ray to NIR SED}\label{sec:appendix}
\begin{figure}
\centering
\includegraphics[width=0.49\textwidth]{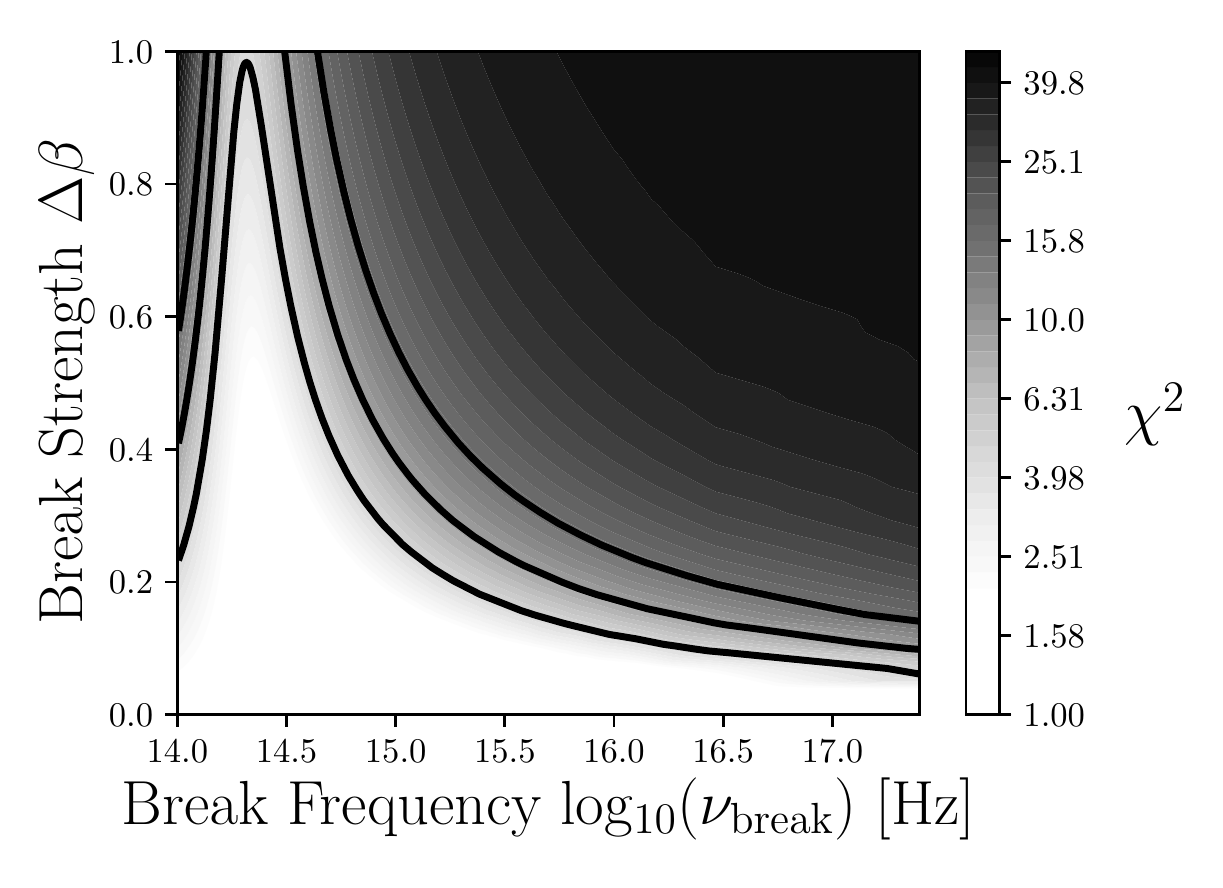}
\caption{The result of fitting a broken power law to the $J$, $H$, $K$ and X-ray fluxes through ${\chi}^{2}$ minimisation. One, two and three ${\sigma}$ contours are shown \citep{1976ApJ...210..642A}. The power law models are fixed to the X-ray data, but we allow for a range of break frequencies and strengths at lower energies. A break in the NIR is favoured, and the possibility that the NIR and X-ray points lie on the same section of the synchrotron spectrum is not excluded.}
\label{fig:appendix}
\end{figure}

In this appendix we detail the fitting procedure used to determine the best spectral fit to the NIR data, given a temporal and spectral extrapolation from the X-ray, as outlined in section \ref{sec:xray}. A broken power law model is used, extrapolated from the X-ray using the {\it Swift}/XRT spectral slope, until a break frequency ${\nu}_\mathrm{break}$ is reached \citep{1998ApJ...497L..17S}. At this break frequency, the spectral slope shallows by an amount ${\Delta}{\beta}$, allowed to vary between 0 and 1, covering a representative range \citep[this break normally occurs between the X-ray and optical, see e.g. ][]{2011A&A...526A..30G}. 
We fit a broken power law from the X-ray to the $J$, $H$ and $K$ points, covering a range of break frequencies and break strengths. Minimising ${\chi}^{2}$ over this parameter space produces best fit values of log$_\mathrm{10}$(${\nu}_\mathrm{break}$)$=14.24$, between the $K$ and $H$ bands, and a strength ${\Delta}{\beta}=0.73$. The range of statistically acceptable fits within 67, 95 and 99.5 per cent confidence regions is shown in Figure \ref{fig:appendix}. 

The flux decrement between the (time-extrapolated) $J$ and the $r$-band is independent of whether a simple ${\Delta}{\beta}=0$ spectral extrapolation is used (which is consistent within the uncertainties), or if the best fit from this procedure is used (where a break is included longwards of this filter). This suggests that the $r$-band non-detection is due to dust or the Lyman-break at high redshift, rather than an afterglow-related spectral break.

\bsp	
\label{lastpage}
\end{document}